\documentclass{article}
\providecommand{\keywords}[1]{\textbf{\textit{Keywords---}} #1}
\usepackage{epsfig}
\usepackage{float}
\usepackage[margin=8.5mm]{caption}
\usepackage[letterpaper, total={7.5in, 9in}]{geometry}

\begin{document}

\title{Validation of the Quantum Physics Processes Underlying the Integrated Optimization of Proton FLASH Radiotherapy}
\author{Nathan Harrison, Serdar Charyyev, Cristina Oancea, \\ Alexander Stanforth, Shuang Zhou, William Dynan, \\ Tiezhi Zhang, Steven Biegalski, Liyong Lin}
\maketitle

\abstract{
\textbf{Purpose:} FLASH is a new treatment modality that requires optimization of dose, dose rate, and linear energy transfer (LET).
In this work we demonstrate the validation of these three quantities under FLASH conditions, which includes the quantum uncertainty in the time-dependent instantaneous dose rate (IDR) curves and LET spectra that underlie the newly proposed integrated optimization framework.

\textbf{Methods:} Measurements of dose, IDR, and LET have been performed at the Emory Proton Therapy Center using a FLASH proton pencil beam with a nominal energy of 250 MeV and a 3D printed ridge filter.
The ridge filter used was designed to uniformly irradiate a spherical target within a water phantom.
Because 3D printing resin is made from an unknown proprietary chemical formula, we developed a method for realistically characterizing and modeling the material in simulations.
Absolute dose in 3D space was measured using a commercial 2D MatriXX PT detector as well as by a novel 4D multi-layer strip ionization chamber (MLSIC), which also simultaneously measures IDR.
Further timing data was measured in the secondary beam by detecting prompt gammas using a Minipix Timepix3; a second detector, Advapix Timepix3, was used to measure LET.
To account for the quantum mechanical nature of particle transport, we developed a technique for detecting individual protons within a high flux primary beam, which was necessary for properly measuring LET spectra.

\textbf{Results:} TOPAS simulations were performed for comparison and showed good agreement with the data, with absolute dose typically having a gamma passing rate of at least 95\% (3 mm/3\% criteria).
Likewise, IDR and LET showed good agreement, with averaged IDR values agreeing within 0.3\% with fluctuations on the order of 10\%, and LET distributions overlapping by at least 85\% and showing an increase in high LET components (greater than 4 $keV/\mu m$) with increasing depth.

\textbf{Conclusion:} As LET and FLASH optimization continues to grow in popularity, measuring IDR, LET, and dose will become even more important, and we expect that the methods described here will prove to be useful tools in radiotherapy treatment planning and QA.
}


\section{Introduction}
\label{sec:Introduction}

Over the past decade, there has been a burgeoning interest in FLASH radiotherapy, where FLASH refers to very high dose rates, typically above 40 Gy/s \cite{kim2022}\cite{diffenderfer2022}.
This interest is due to a number of studies \cite{favaudon2014}\cite{loo2017}\cite{vozenin2019}\cite{montaygruel2017} which demonstrated that these high dose rates can significantly reduce damage to the healthy tissue of organs at risk (OARs) during treatment when compared to more conventional techniques.
Proton pencil beam scanning (PBS) systems are an especially promising candidate for delivering FLASH treatments since many existing proton therapy centers can be made capable of delivering such beams with minimal overhead \cite{zou2021}\cite{kang2022}.
Besides FLASH dose rates, which typically deliver a dose within a few milliseconds, protons offer biological effectiveness related to linear energy transfer (LET) according to their spatial and timing distributions,
i.e. quantum physics processes besides the traditional classical physics processes that are described by dose distributions \cite{paganetti2012}\cite{paganetti2019}.

One approach to treating patients with a FLASH proton (or other charged particle) beam is to use a patient-specific ridge filter to modulate the beam and therefore deliver an optimal dose distribution within a given target volume \cite{simeonov2017}\cite{lin2015b}.
Advancements in 3D printing technology make fabricating these patient-specific ridge filters accessible and affordable, and 3D printing has already been shown to be a useful tool in radiotherapy applications \cite{simeonov2017}\cite{mayer2015}.
Recently, work has been done to show that in addition to dose, the dose rate and LET distributions can all simultaneously be optimized using a ridge filter \cite{rliu2022}.

All of these new developments have generated an increased demand for innovation in detectors and techniques for measuring dose, dose rate, and LET.
In this work we use four detectors for validating dose, dose rate, and LET with a FLASH proton pencil beam and a ridge filter:
\begin{enumerate}
  \item A commercial moving DigiPhant+MatriXX PT detector for measuring the 3D absolute dose distribution \cite{lin2015}.
  \item A 2D Advapix Timepix3 detector from Advacam, which is a pixelated silicon detector with a 14 mm $\times$ 14 mm sensitive region with 256 $\times$ 256 pixels and nanosecond scale timing resolution, for measuring the 3D LET distribution \cite{llopart2007}.
  \item A 2D Minipix Timepix3 detector from Advacam, with similar specs as the Advapix Timepix3, for measuring timing \cite{granja2018} via prompt gamma rays \cite{oancea2022}\cite{charyyev2022}.
  \item A MLSIC 4D absolute dose and timing detector \cite{zhou2022}, which is a novel detector and was used for the first time with FLASH in this work. Such strip ionization chamber detectors are growing in popularity, for example \cite{yang2022} describes a related study.
\end{enumerate}
The data from these detectors was then compared to GEANT based TOPAS simulations \cite{perl2012}.

The techniques for measuring dose, dose rate, and LET presented here can be used to validate physical and biological optimization frameworks, such as, for example, the ones presented in \cite{cliu2020} and \cite{shan2018}.
This manuscript represents the first validation of quantum physics key parameters under primary FLASH beams.

\section{Materials and Methods}
\label{sec:Methods}

\begin{figure}[H]
\centering
\includegraphics[width=5in]{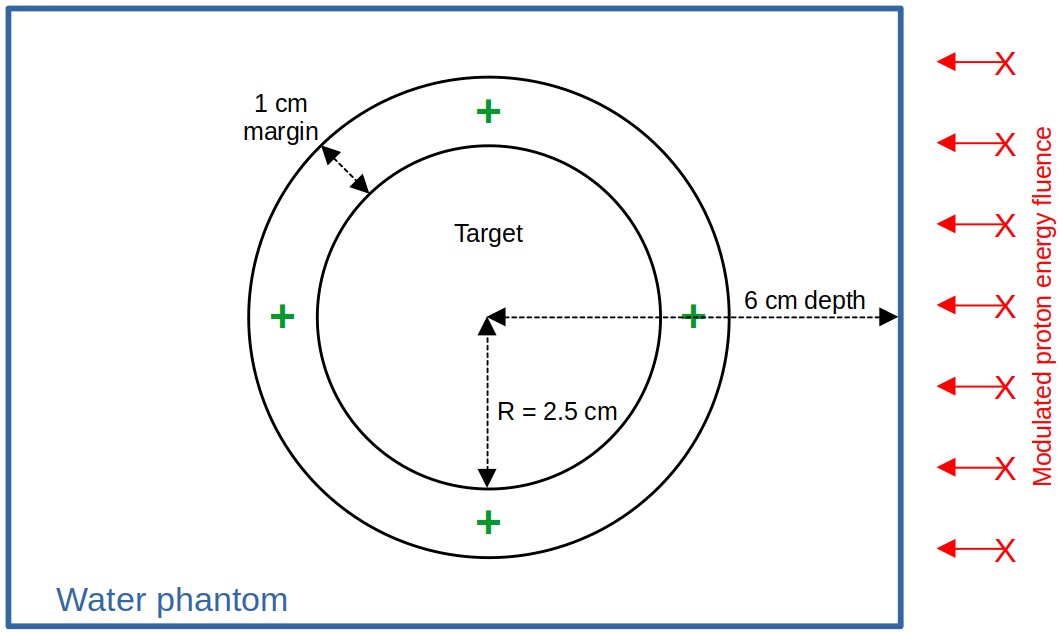}
\includegraphics[width=5in]{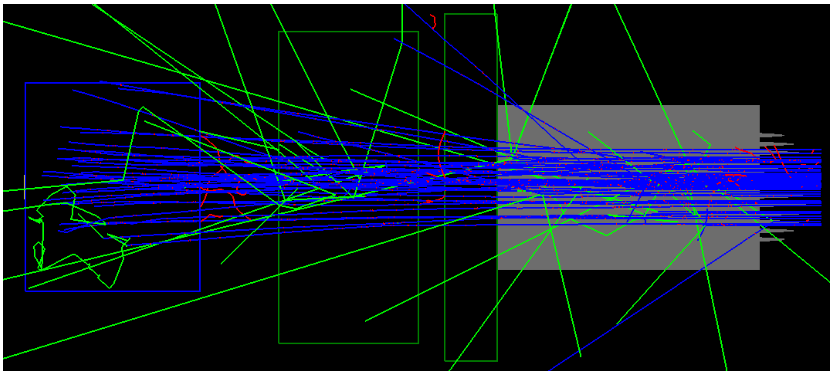}
\caption{Top: Water phantom with a 50 mm diameter spherical target and 10 mm margin at a depth of 60 mm. A ridge filter is used to create the desired proton energy fluence.
The green ``+'' symbols highlight the proximal, distal, and lateral margins, which are regions of particular importance.
Bottom: A Monte Carlo simulation of (from right to left) the ridge filter, a 30 mm range shifter, the 80 mm PMMA block, and water phantom.
Also shown by blue lines are 250 MeV protons travelling from right to left along with secondary particles shown in red and green.}
\label{fig:problem_definition}
\end{figure}

The purpose of a ridge filter is to create a desirable proton energy fluence from a monoenergetic beam, as shown in Figure~\ref{fig:problem_definition}.
This fluence then determines the dose, dose rate, and LET in the target and surrounding regions.
The margin is of particular interest since it is simultaneously in close proximity to the target and OARs.

\begin{figure}[H]
\centering
\includegraphics[width=6.5in]{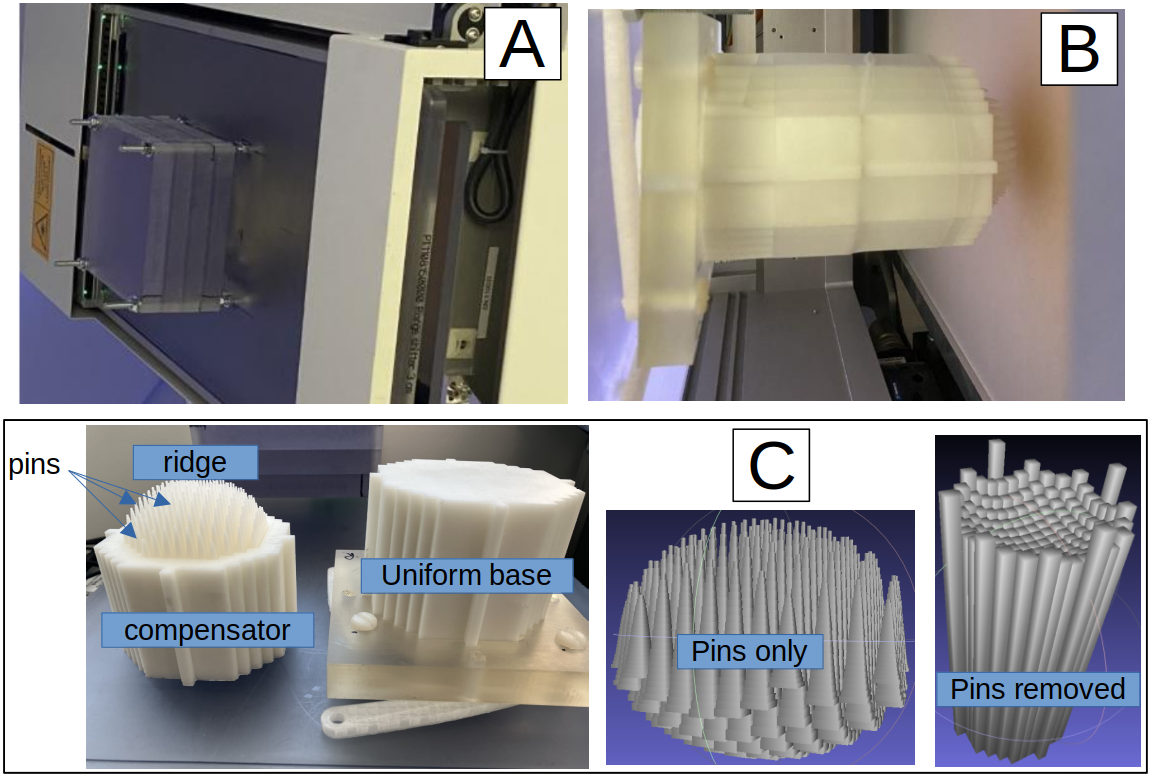}
\caption{A: The outer (downstream) part of the nozzle. B: The inner (upstream) part of the nozzle where the ridge filter assembly is mounted.
C: Photos and CAD images of the components of the ridge filter, which include a reuseable uniform base, a compensator, and pins.}
\label{fig:ridge_anatomy}
\end{figure}

The design of the ridge filter was largely base on Simeonov et al \cite{simeonov2017} and is meant to uniformly irradiate a 70 mm diameter spherical region inside of a water phantom with the target center at a depth of 60 mm.
Figure~\ref{fig:ridge_anatomy} shows the different components of the ridge filter, which was printed in two halves (uniform base and compensator+ridge) so the uniform base could be reused.

\subsection{Calibration}
\label{subsec:Calibration}

\subsubsection{Material Characterization}
\label{subsubsec:MaterialCharacterization}

The ridge filter was 3D printed using a proprietary resin from Formlabs called Rigid 4000.
Since the exact chemical formula of the resin is not public information, we had to develop a technique for realistically characterizing and modeling the material for simulations.
To do this, an 80 mm long and 20 mm diamter cylinder was printed with the same resin.
We then used a proton beam and commercial Zebra detector from IBA to measure the R80 of the protons with and without the cylinder in the beam.
The water equivalent thickness (WET) of the cylinder is then
\begin{equation}
\label{eq:wet}
WET = R80_{without} - R80_{with}.
\end{equation}
From here, the relative stopping power (RSP) can be calculated by 
\begin{equation}
\label{eq:rsp}
RSP = \frac{WET}{material\ length}.
\end{equation}
To realistically model the stopping power of this material in the simulations, we assumed a density $\rho_{resin} = (RSP)\rho_{water}$.
To realistically model the scattering power, we assumed a chemical composition of $xH_2O + (1-x)SiO_2$, and then simulated many values of $x$ between 0 and 1 and compared the dose profile to data to choose the best value for $x$.
$SiO_2$ was chosen because it is known to be one of the major ingredients in resins besides PMMA \cite{zakeri2020}.

\subsubsection{LET Measurements}
\label{subsubsec:LET Measurements}

\begin{figure}[H]
\centering
\includegraphics[width=4.0in]{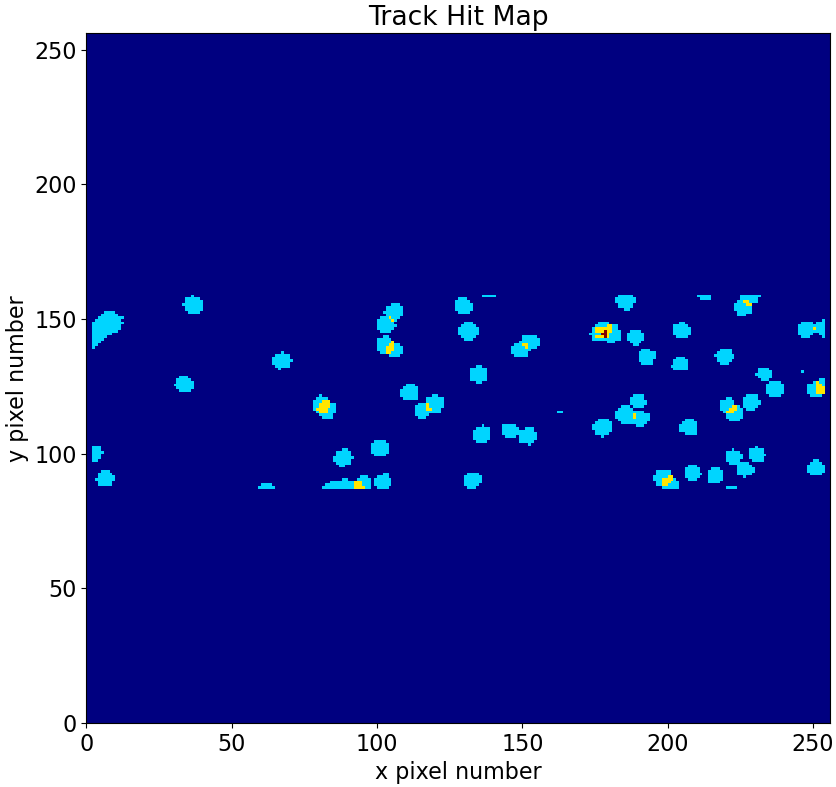}
\caption{Advapix Timepix3 detector response for a single frame with roughly 50 particle tracks. All 256$\times$256 pixels (14 mm $\times$ 14 mm) are shown, but the effect of the two-thirds mask, which turns off two-thirds of the pixels, can clearly be seen at the top and bottom.}
\label{fig:detector_mask}
\end{figure}

For accurate LET measurements, the Advapix Timepix3 manufacturer recommends setting the detector acquisition time to a value no less than 500 $\mu s$ and to orient the detector such that the beam impinges upon the sensor at a $45^\circ$ angle.
However, due to minimum current constraints of our beam, the detector became saturated under these conditions and it was therefore necessary to use shorter acquisition times and to orient the detector perpendicular to the beam to produce smaller clusters (i.e. fewer pixels).
The shortened acquisition time lead to under-counting the deposited energy due to incomplete charge collection by the detector electronics.
The speed of the electronics was improved by ``masking'' two-thirds of detector, i.e. essentially turning off two-thirds of the pixels, as shown in Figure~\ref{fig:detector_mask}, which shows the 256$\times$256 pixel (14 mm $\times$ 14 mm) arrangement with the top third and bottom third masked.
The perpendicular detector orientation also lead to an underestimate of the LET, an effect described in \cite{granja2018b}, \cite{granja2021}, and \cite{nabha2022}.

To correct for these effects, we performed systematic studies of the detector response as a function of both acquisition time and detector angle in order to come up with LET correction factors to apply to the collected experimental data.
This was done by measuring LET distributions with different acquisition times (or detector angles), fitting each with a Gaussian to find the peak location, and then plotting the peak position as a function of acquisition time (or detector angle).

\subsection{Experimental Setup}
\label{subsec:ExperimentalSetup}

\begin{figure}[H]
\centering
\includegraphics[width=6.5in]{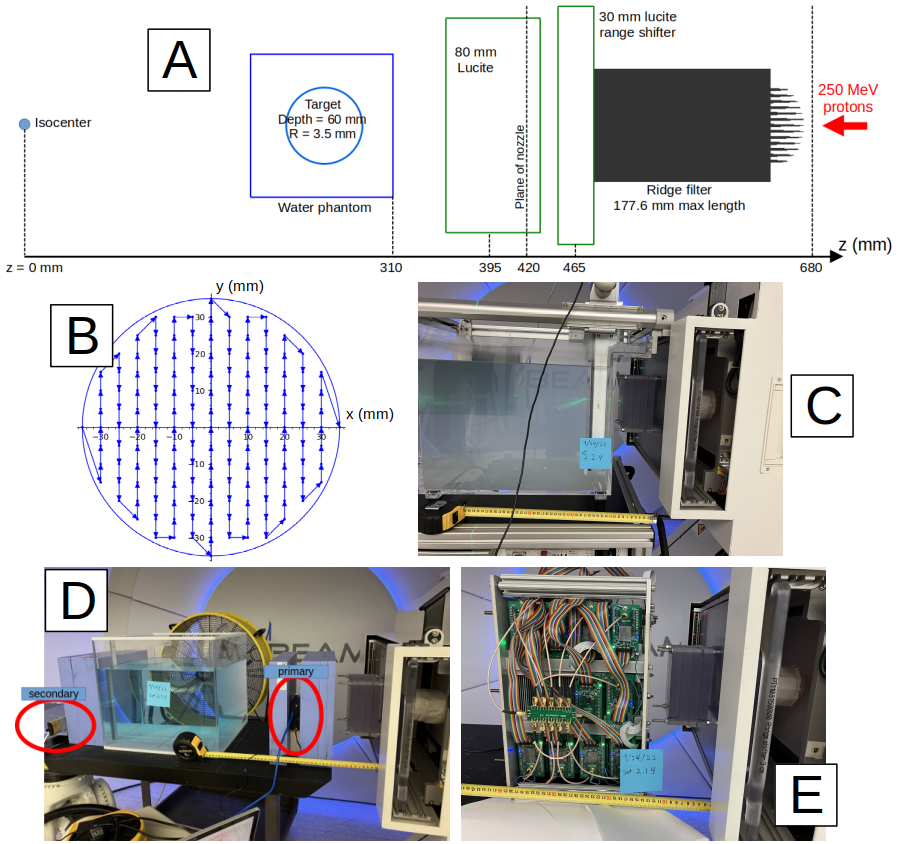}
\caption{(A) The experimental setup showing the ridge filter, 30 mm lucite range shifter, 80 mm block of additional lucite, and water phantom.
(B) The spot map for the proton pencil beam.
(C) A photo of the setup for the MatriXX PT absolute dose measurement.
(D) A photo the setup for the LET and timing measurements done with the two Timepix3 detectors.
(D) A photo the setup for the dose and dose rate measurements done with the MLSIC detector.}
\label{fig:experimental_setup}
\end{figure}

Figure~\ref{fig:experimental_setup}A shows the experimental setup, where 250 MeV protons first impinge upon the ridge filter, and then pass through an additional 30 mm + 80 mm of lucite in order to modulate the protons to the desired depth.
From there, the protons deposit their remaining energy in the water phantom.
Also shown is the spot map (Figure~\ref{fig:experimental_setup}B) for the proton pencil beam, which consists of 149 spots with 5 mm spacing.

The nozzle of the machine is equipped with a laser grid running parallel to the downstream face of the range shifter for safety purposes, such that the beam will be shut off if any of the lasers are blocked.
The mounting mechanism of the 80 mm lucite block had to be carefully designed to avoid blocking these lasers.
This was done by mounting the block via four narrow bolts that could fit in-between adjacent lasers.
The bolts allowed the block to be mounted such that there is a 1.50 mm air gap between the 30 mm range shifter and 80 mm block, thus avoiding blocking the lasers.

Figure~\ref{fig:experimental_setup}C-E show photographs of the experimental setups with the DigiPhant+MatriXX PT, Timepix3 detectors, and the MLSIC detector, respectively.
Aside from changing detectors, no other changes were made to the setup for these three different measurements.

Since the sensor of the Advapix Timepix3 detector used to measure LET is rather small (14 mm $\times$ 14 mm) (and made even smaller by masking), each LET measurement consisted of three ``sub-measurements'', where each sub-measurement was done with the detector at a different lateral position.
The three lateral offsets were 0 mm, 14 mm, and 28 mm.
The three sub-measurements were then patched together using a technique similar to the pair magnification method described in \cite{lin2013}.

\subsection{Measurement Details and Workflow}
\label{subsec:Workflow}

The proton PBS system used was the Varian ProBeam, which can deliver energies up to 250 MeV at nozzle currents beyond 300 nA with the latest monitor unit chamber, although we were limited to 100 nA by our current monitor unit chamber.
Each of the 149 spots received 250 monitor units (MU), where a MU is proportional to the number protons ($N_p$).
The MU to $N_p$ conversion factor is energy dependent; at 250 MeV, there are $5.343 \times 10^6$ protons per MU \cite{chang2020}\cite{charyyev2022b}.

Three datasets were collected, one for each of the detector configurations described in Figure~\ref{fig:experimental_setup}.
Set 1 was the MLSIC measurement; since the MLSIC is a 4D detector, all the data was collected by running the beam one single time.
Set 2 was the DigiPhant+MatriXX PT measurement; absolute dose in the xy-plane was measured at 17 different depths between 28 mm and 95 mm.
Set 3 was the measurement with two Timepix3 detectors; the primary (upstream) detector was used to measure LET while the secondary (downstream) detector was used to measure timing.
Set 3 consisted of several runs for different depths and lateral offsets of the primary detector.

\section{Results}
\label{sec:Results}

\subsection{Material Characterization and Dose with 2D Detector}
\label{subsec:Calibration_Dose2D}

The characterization of the Formlabs Rigid 4000 resin yielded an RSP value of 1.265 and a $x$-value of 0.95.

\begin{figure}[H]
\centering
\includegraphics[width=7in]{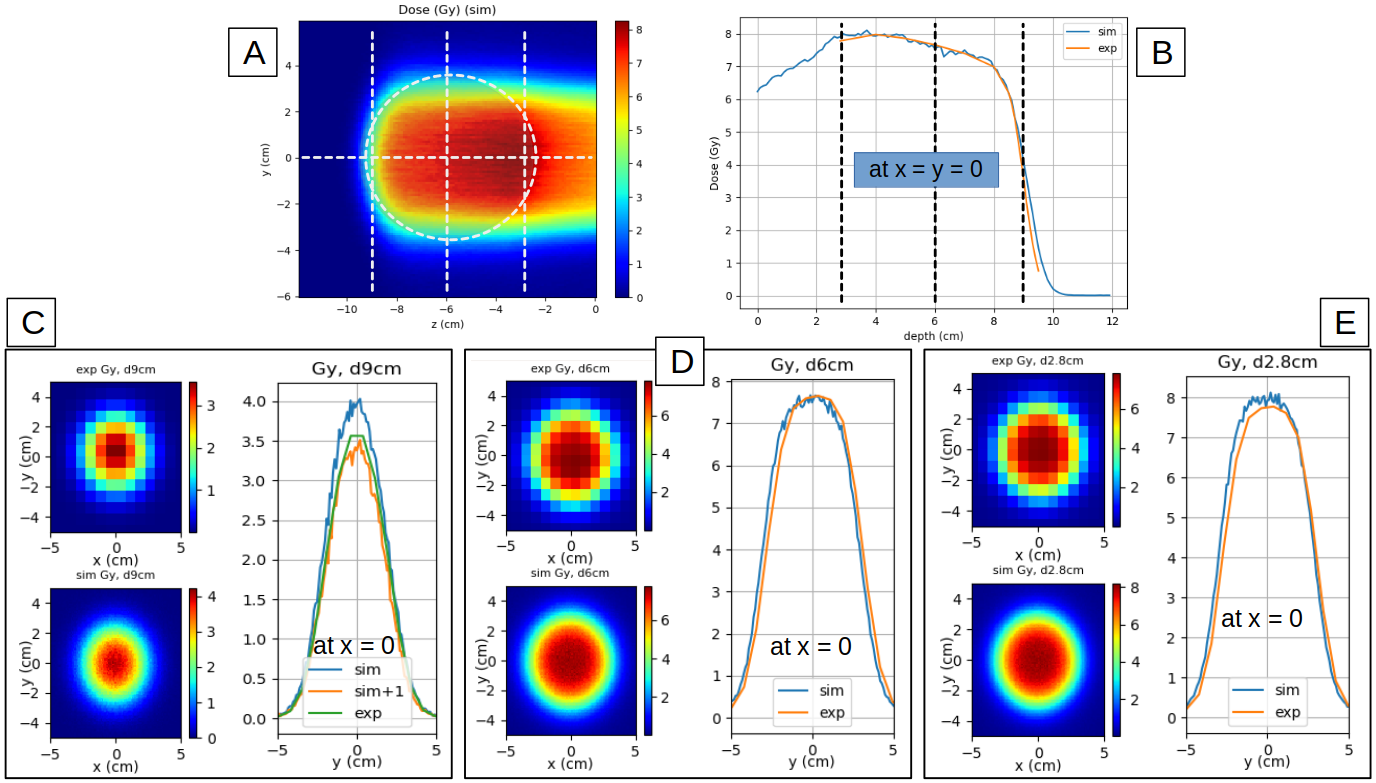}
\caption{A: Sagittal slice at the mid-plane ($x=0$) of absolute dose from the Monte Carlo simulation;
the dashed lines represent the spherical target, the central axis, and three representative depths which are shown in more detail in (C)-(E).
B: Depth dose along the central axis ($x=y=0$) for MatriXX PT data (orange) and simulation (blue); the dashed lines represent the three depths which are shown in more detail in (C)-(E).
C-E: Absolute dose distributions at depths of 90 mm, 60 mm, and 28 mm, respectively for data and simulations; in (C), an additional simulation result is shown at depth 91 mm to show that the disagreement is within 1 mm.}
\label{fig:abs_dose_matrixx_sim}
\end{figure}

Figure~\ref{fig:abs_dose_matrixx_sim} shows the absolute dose measured with the MatriXX PT along with simulations for comparision.
In Figure~\ref{fig:abs_dose_matrixx_sim}C, which shows data at depth 90 mm, which is within the distal falloff region, an additional simulation result at depth 91 mm is shown to demonstrate that the 10\% disagreement between data and simulation represents a less than 1 mm difference.
A gamma analysis was performed between the dose measured with the MatriXX PT and TOPAS simulation.
A 3D gamma analysis was performed for each measurement using in-house software following the algorithm described in \cite{ju2008}.
The gamma analysis was limited to points greater than 5\% the maximum dose of each measurement.
Using a 3\%/3mm criteria, 16 of the 17 measurements had a passing rate greater than 90\% (representing depths from 28 mm to 94 mm).
The measurement at a depth of 95 mm was the only measurement with a gamma passing rate less than 90\%, though the maximum dose in this measurement was 0.751 Gy, representing a dose less than 10\% the maximum dose of the plan.
Of the 16 measurements which had at least a 90\% passing rate, 15 passed with at least 95\% of points meeting the 3\%/3mm criteria (representing depths from 28 mm to 94 mm).

\subsection{Dose and Dose Rate with MLSIC}
\label{subsec:Dose_DR_MLSIC}

\begin{figure}[H]
\centering
\includegraphics[width=7in]{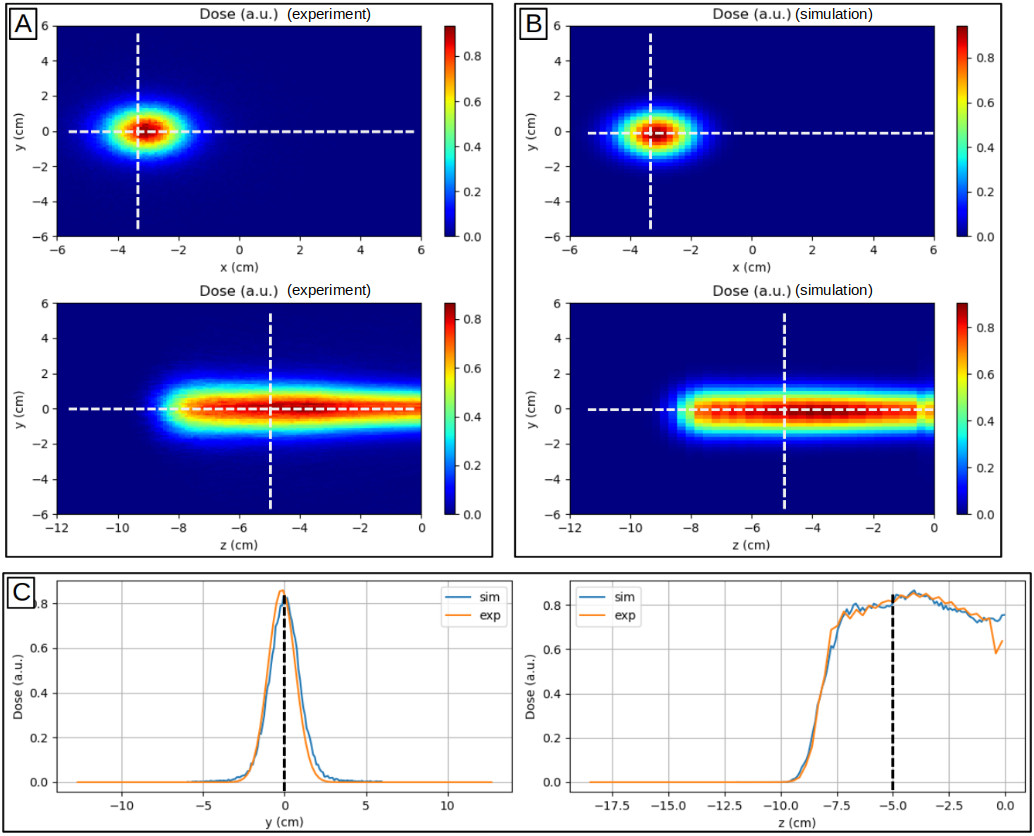}
\caption{Dose distributions for the first spot.
A: Axial slice at depth 50 mm (top) and sagittal slice at $x=0$ (bottom) from simulation.
B: Axial slice at depth 50 mm (top) and sagittal slice at $x=0$ (bottom) from MLSIC data.
C: Relative dose profile at depth 50 mm for MLSIC data (orange) and simulation (blue) (left) and relative depth dose curve for MLSIC data (orange) and simulation (blue).}

\label{fig:mlsic_results}
\end{figure}

The MLSIC detector is comprised of $x$ and $y$ strips at different depths and principally reconstructs the 3D dose and dose rate distributions with certain assumptions about the dose profiles of the pencil beams \cite{zhou2023}.
The presence of the ridge filter causes irregular dose profiles and makes dose reconstruction from MLSIC data challenging, we therefore demonstrate the results of one selected spot, specifically the first spot of the spot map.
This data can be seen in Figure~\ref{fig:mlsic_results}.

The measurements done with the MLSIC detector used a 7 nA beam with 149 spots and 250 MU per spot.
The irradiation time (IRT) measured by MLSIC was 26.928 ms.
This should be compared to the value from Varian log files which recorded an IRT of 27.265 ms, a 1.2\% difference.

\begin{figure}[H]
\centering
\includegraphics[width=4in]{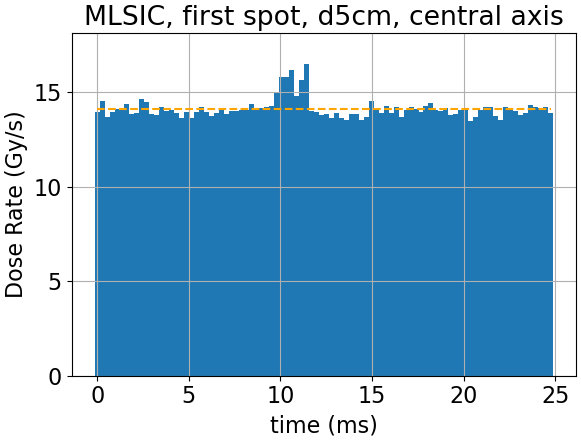}
\caption{Time-dependent instantaneous dose rate fluctuations of the 99 samples for spot 1 measured with the MLSIC detector at a depth of 50 mm along the central axis of the spot (blue solid line).
The orange dashed line represents the expected dose rate according to simulations.}
\label{fig:mlsic_doserate}
\end{figure}

The time-dependent instantaneous dose rate curve for the first spot was also measured.
The integration duration of the MLSIC is 272 $\mu s$, which corresponds to 99 samples within the time window of the first spot.
The relative dose per sample, which remained fairly constant but for fluctuations on the order of 10\% or less, was scaled to absolute dose using simulation data.
This was then used to calculate dose rate, which is plotted in Figure~\ref{fig:mlsic_doserate} at a depth of 50 mm along the central axis of the spot.
The small variations in dose rate are most likely due to quantum fluctuations in the anode, cathode, and RF of the cyclotron \cite{charyyev2022b}.
For this data, the average number of protons within each 272 $\mu s$ sample is $1.349 \times 10^7$ and the peak at 11.5 ms corresponds to $1.581 \times 10^7$ protons.

\subsection{Timing with Secondary Pixelated Detector}
\label{subsec:Timing_minipix}

Dose rate is calculated by simply dividing the total dose by the IRT.
Since the dose measurements have already been completed as described above, dose rate measurements can be simplified to just measuring the IRT.

\begin{figure}[H]
\centering
\includegraphics[width=3.5in]{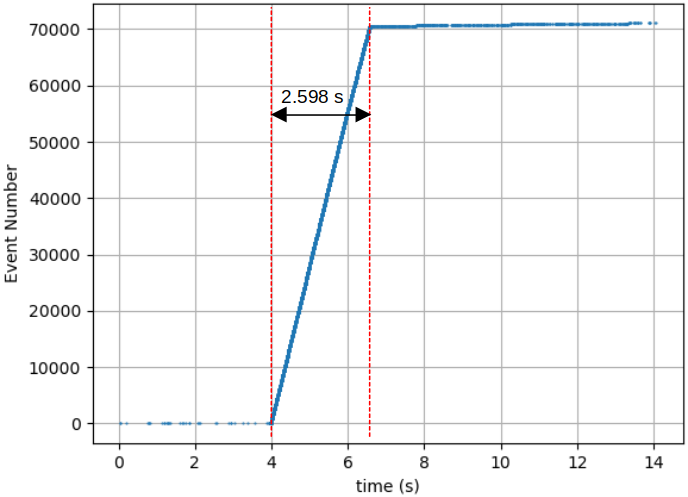}
\includegraphics[width=3.5in]{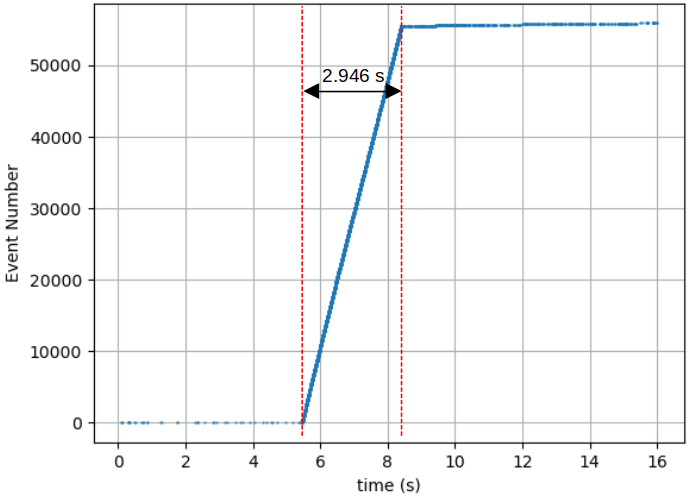}
\includegraphics[width=3.5in]{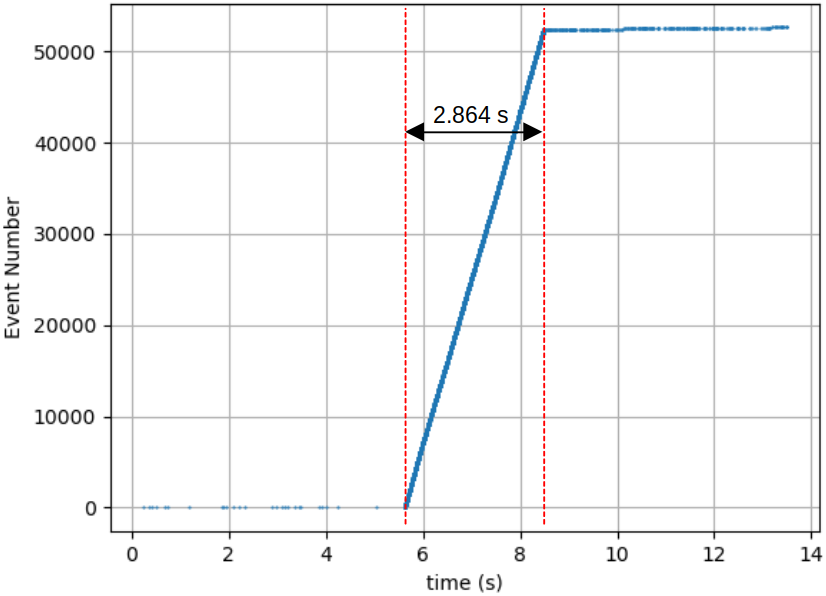}
\includegraphics[width=3.5in]{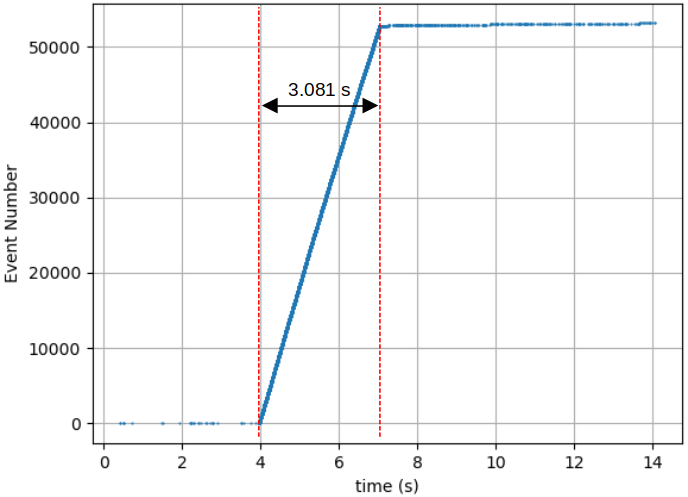}
\caption{Timing with secondary detector (Minipix Timepix3) at 250 MeV and 10 nA. Four repetitions were done to demonstrate reproducibility.}
\label{fig:mpx_dt}
\end{figure}

Figure~\ref{fig:mpx_dt} shows timing measured with Minipix Timepix3 at 250 MeV, 10 nA, 250 MU per spot, and 149 spots.
Four repetitions were done to demonstrate reproducibility.
The plots show clearly when the beam first turns on and when it stops, making extraction of the timing information from the data very straightforward, in this case between 2.598 and 3.081 seconds.
Varian log files recorded IRTs ranging from 2.58553 to 3.06717 seconds, with an average of 2.86393 seconds, showing good agreement with the detector within 0.3\%.

Commissioning data from Raystation reports that at 250 MeV, there are $5.343 \times 10^6$ protons per MU.
Therefore, the expected IRT for these measurements is

\begin{equation}
\label{eq:irt}
\frac{(5.343 \times 10^6 \ prot/MU) \times (250 \ MU/spot) \times (149 \ spots) \times (1.6 \times 10^{-19} \ C/prot)}{(10 \times 10^{-9} \ C/second)}
\end{equation}

or 3.184428 seconds.

\subsection{LET Calibration and Results with Primary Pixelated Detector}
\label{subsec:LET_advapix}

\begin{figure}[H]
\centering
\includegraphics[width=7in]{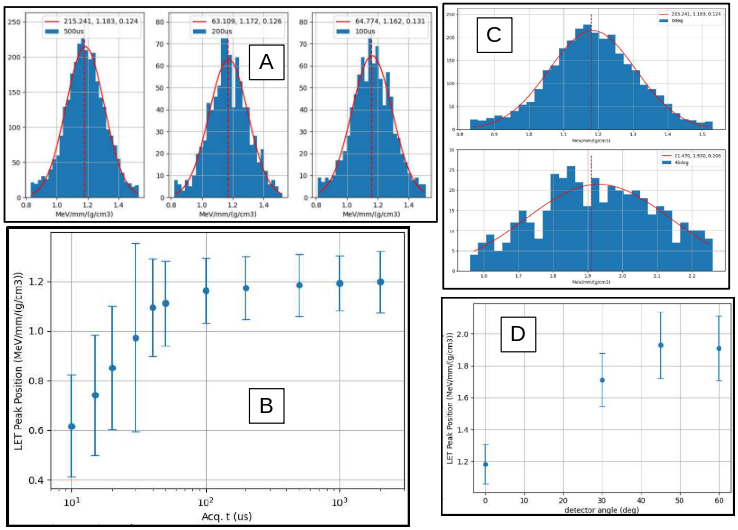}
\caption{A: Three representitive LET distributions with different acquisition times (500, 200, and 100 $\mu s$, respectively) with Gaussian fits used to find the peak position.
B: LET peak position vs acquisition time using results from (A).
C: Two representitive LET distributions with different detector angles ($0^\circ$ and $45^\circ$, respectively) with Gaussian fits used to find the peak position.
D: LET peak position vs detector angle using results from (C).}
\label{fig:calibration_plots}
\end{figure}

For the LET correction factors, each LET distribution was fit with a Gaussian in order to find the location of the peak, as shown in Figures~\ref{fig:calibration_plots}A and \ref{fig:calibration_plots}B.
The peak positions were then plotted as a function of acquisition time (Figure~\ref{fig:calibration_plots}B) and detector angle (Figure~\ref{fig:calibration_plots}D) which can then be used to calculate the necessary LET correction factors.

\begin{figure}[H]
\centering
\includegraphics[width=7in]{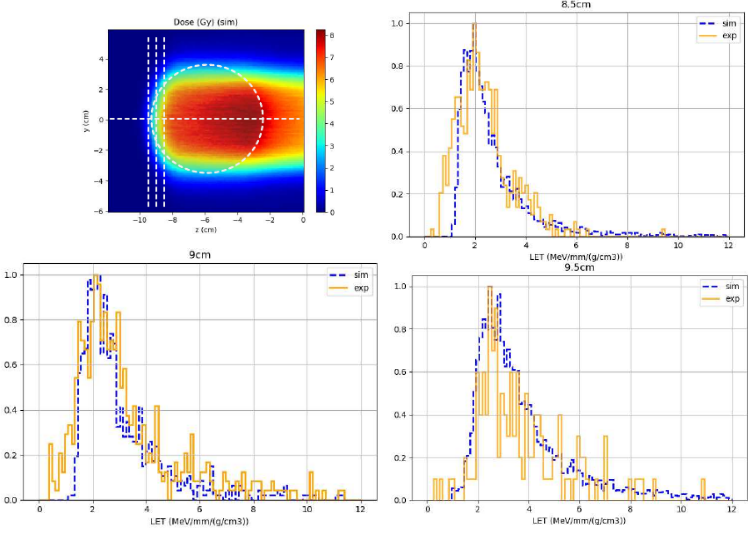}
\caption{Top left: Sagittal view of dose distribution with dashed lines showing the spherical target, the central axis, and the three depths where the LET was measured.
The three remaining plots show the LET distributions for simulation (blue dashed line) and Advapix Timepix3 data (solid orange line) at depths of 85 mm (top right), 90 mm (bottom left), and 95 mm (bottom right).}
\label{fig:let}
\end{figure}

Measurements of the LET distributions were done using the Advapix Timepix3 detector and requires some filtering of the data to remove noise and background.
Details on filtering can be found in \cite{granja2021} and \cite{charyyev2021}.
Figure~\ref{fig:let} shows the corrected experimental and simulated LET distributions for three depths: 85, 90, and 95 mm; the overlaps between the experimental and simulated distributions were 85.6\%, 89.9\%, and 85.3\%, respectively.
The data show more and more high LET (greater than 4 $keV/\mu m$) components with increasing depth.
Despite using very short acquisition times and a perpendicular detector orientation, measurements shallower than 85 mm were not possible due to detector saturation.

\section{Discussion}
\label{sec:Discussion}

There have been several optimization methods to include FLASH dose rate and/or LET. Here we use time-dependent instant dose rate curves and LET spectra to describe the quantum physics processes underlying such integrated optimizations.
This manuscript represents the first validation of quantum physics key parameters under unmodified primary FLASH beams.
This is possible due to our novel method using a combination of masking and microsecond acquisition time to avoid the saturation and recover the original LET values by calibration of under response of such short acquisitions.

Although it was only possible to measure LET at three depths near the distal side of the target, what we present here demonstrates a proof of concept.
As FLASH radiotherapy continues to grow in popularity, advancements in detector technology will be important.
Advacam is currently working on a detector with a faster chip that will likely solve some of the struggles we experienced with measuring LET.
Similarly, our measurements with the MLSIC detector demonstrate a proof of concept for measuring the 4D dose distribution, and the WUSTL group is also currently working on a new and improved version of the MLSIC.

In Figure~\ref{fig:mlsic_results} we show a classical physics representation of a spot dose distribution.
This concept of a spot is invalid under FLASH, which in reality is represented by quantum physics uncertainty of positioning, which can be seen over time by Figures~\ref{fig:detector_mask} and \ref{fig:mlsic_doserate} with uncertainty of proton energy represented by LET spectra in Figure~\ref{fig:let}.

Although this work was done with a ridge filter and proton beam, we believe that the methods described have broad applicability to non-ridge and non-proton therapy modalities as well.
For example, \cite{qliu2013} and \cite{okamoto2011} both describe related work with photon beams.
Furthermore, with Liu et al \cite{rliu2022} recently showing the fesibility of simultaneously optimizing dose, dose rate, and LET, our techniques will be important for validating solutions to that optimization problem, as it is necessary to validate the underlying quantum physics processes related to FLASH dose rate and LET.

\section{Conclusion}
\label{sec:Conclusion}

In this manuscript we present methods for obtaining dose, dose rate, LET, and proton energy fluence with a FLASH pencil beam and ridge filter, along with experimental and GEANT4 based simulated data.
Despite some detector limitations, our measured data typically agreed with simulations within a few percent.
We have also developed a technique for characterizing and modeling unknown materials for particle transport simulation purposes.
Additionally, we have developed a method for detecting individual protons within a high-flux beam, which is important for LET measurements given the quantum mechanical nature of LET.
We expect that the methods described here will prove to be useful tools in radiotherapy treatment planning.

\section{Acknowledgment}
\label{sec:Acknowledgment}

We would like to thank Georgia Institute of Technology students Hugh Nunnally and Caleb Corliss for their help with 3D printing and ridge mount design as well as Varian engineer Phillip Kelly for sharing his extensive knowledge of the Varian ProBeam machine and fruitful discussions about experimental design.
We would also like to thank Ruirui Liu for helpful discussions on ridge filter design and manufacturing.
The MLSIC portion of this work was made possible by NCI SBIR Contract 75N91022C00055 with PI Tiezhi Zhang.
Additional funding was provided by Liyong Lin's Emory University faculty startup fund.
Finally, we are grateful to Katja Langen and Roelf Slopsema of Emory University for providing us with beam time at the Emory Proton Therapy Center.

\end{document}